\documentclass[iop]{emulateapj}
\usepackage{epsfig} 
\usepackage{mathrsfs,amssymb}
\usepackage{graphicx,latexsym}
\usepackage{verbatim}

\newcommand{\Ha}{H$\alpha$}
\newcommand{\Sigd}{$\Sigma_d$}

\newcommand{\hii}{{\sc Hii}}
\newcommand{\hi}{{\sc Hi}}
\newcommand{\nhi}{$N$({\sc Hi})}
\newcommand{\oiii}{{\sc Oiii}}
\newcommand{\sii}{{\sc Sii}}

\newcommand{\mic}{$\mu$m}

\def\spose#1{\hbox to 0pt{#1\hss}}
\def\dt{\spose{\raise 1.0ex\hbox{\hskip2pt$\mathchar"201$}}}    

\slugcomment{*** Accepted May 15, 2017 ***}

\shorttitle{\hii\ regions and 8 \mic\ and 24 \mic\ emission}
\shortauthors{Oey et al.}

\begin{document}

\title{Dust emission at 8~\mic\ and 24~\mic\ as Diagnostics of \hii\ Region Radiative Transfer}

\author{M. S. Oey\altaffilmark{1}, J. L\'{o}pez-Hern\'{a}ndez\altaffilmark{1,10}, J. A. Kellar\altaffilmark{1,11}, E. W. Pellegrini\altaffilmark{2}, 
K. D. Gordon\altaffilmark{3}, K. E. Jameson\altaffilmark{4}, A. Li\altaffilmark{5},
S. C. Madden\altaffilmark{6}, M. Meixner\altaffilmark{3}, J. Roman-Duval\altaffilmark{3},
C. Bot\altaffilmark{7}, M. Rubio\altaffilmark{8}, and A. G. G. M. Tielens\altaffilmark{9}}
\altaffiltext{1}{Department of Astronomy, University of Michigan, 311 West Hall, 1085 S. University Ave. Ann Arbor, MI, 48109-1107}
\altaffiltext{2}{Institut f\"ur Theoretische Astrophysik,
  Albert-\"Uberle-Str. 2, D-69120 Heidelberg, Germany}
\altaffiltext{3}{Space Telescope Science Institute, 3700 San Martin Dr, Baltimore, MD 21218}
\altaffiltext{4}{Astronomy Department and Laboratory for
  Millimeter-wave Astronomy, University of Maryland, College Park, MD 20742}
\altaffiltext{5}{Department of Physics and Astronomy, University of
  Missouri, Columbia, MO 65211}
\altaffiltext{6}{Laboratoire AIM, CEA, Universit\'e Paris VII,
  IRFU/Service d'Astrophysique, Bat. 709, F-91191 Gif-sur-Yvette, France}
\altaffiltext{7}{Observatoire Astronomique de Strasbourg,
    Universit\'e de Strasbourg, CNRS, UMR 7550, 11 Rue de l'Universit\'e,
    F-67000 Strasbourg, France}
\altaffiltext{8}{Departamento de Astronom\'\i a, Universidad de Chile,
  Casilla 36-D, Santiago, Chile}
\altaffiltext{9}{Leiden Observatory, Leiden University, PO Box 9513,
  NL-2300RA Leiden, The Netherlands}
\altaffiltext{10}{Present address:  Fac. de Ciencias de la Tierra y del Espacio,
  Universidad Autonoma de Sinaloa, Blvd. de las Americas y
  Av. Universitarios S/N, Ciudad Universitaria, C. P. 80010
  Culiac\'an, Mexico}
\altaffiltext{11}{Private address}
  

\begin{abstract}
We use the {\sl Spitzer} SAGE survey of the Magellanic Clouds to
evaluate the relationship between the 8~\mic\ PAH emission, 24~\mic\ hot
dust emission, and \hii\ region radiative transfer.  We confirm that in
the higher-metallicity Large Magellanic Cloud, PAH destruction is sensitive to optically
thin conditions in the nebular Lyman continuum:  objects identified as
optically thin candidates based on nebular ionization structure
show 6 times lower median 8~\mic\ surface brightness (0.18 mJy
arcsec$^{-2}$) than their optically thick counterparts (1.2 mJy
arcsec$^{-2}$).  The 24~\mic\ surface brightness also shows a factor
of 3 offset between the two classes of objects (0.13 vs 0.44 mJy
arcsec$^{-2}$, respectively), which is driven by the association
between the very small dust grains and 
higher density gas found at higher nebular optical depths.  In
contrast, PAH and dust formation in the low-metallicity Small
Magellanic Cloud is strongly inhibited such that we find no variation
in either 8~\mic\ or 24~\mic\ emission between our optically thick and
thin samples.  This is attributable to extremely low PAH and dust production
together with high, corrosive UV photon fluxes in this low-metallicity
environment.  The dust mass surface densities and gas-to-dust ratios
determined from dust maps using {\sl Herschel} HERITAGE survey data
support this interpretation.  
\end{abstract}

\keywords{radiative transfer --- stars: massive --- dust, extinction
  --- \hii\ regions --- galaxies: ISM --- Magellanic Clouds}

\section{Introduction} 

The ionizing radiation from massive stars has fundamental
consequences on scales ranging from individual circumstellar
disks to the ionization state of the entire universe.  On galactic scales,
the escape fraction of Lyman continuum radiation from galaxies 
is crucial to the ionization state of the intergalactic medium 
and cosmic reionization of the early universe; and radiative feedback
is also a major driver for the energetics and phase balance of the
interstellar medium (ISM) in star-forming galaxies.  Thus, determining
the fate of ionizing photons from high-mass stars is critical to
understanding the formation and evolution of galaxies throughout
cosmic time.  

Within star-forming galaxies, it has long been recognized that the diffuse, warm ionized
medium (WIM), which is the most massive component of ionized gas in
galaxies \citep{walterbos98}, is energized by OB stars 
\citep[e.g.][]{haffneretal09}. The WIM is a principal component of the multi-phase
ISM, and strongly prescribes galactic ecology, which drives evolutionary
processes like star formation and galactic dynamics.  The standard paradigm is that
the WIM is powered both by ionizing radiation escaping from classical
\hii\ regions, and by field OB stars \citep[e.g.][]{oeykennicutt97,hoopes00}.  
While additional ionizing
sources are sometimes suggested, it is clear that only massive stars
can provide enough power to generate the WIM \cite[e.g.][]{reynolds84},
although other mechanisms may be secondary contributors.

The relative importance of optically thin \hii\ regions vs field star
ionization of the WIM is still poorly understood.  Comparison of
predicted and observed \hii\ region luminosities in nearby galaxies
had suggested that both sources are not only viable, but necessary
\citep{oeykennicutt97, hoopes00, hoopesetal01}.
However, modern stellar atmosphere models for massive stars
\citep[e.g.][]{martinsetal05,pauldratchetal01}
exhibit lower ionizing fluxes than those of the previous generation,
casting doubt that a significant fraction of classical \hii\ regions are
density-bounded \citep[optically thin;][]{voguesetal08}.  On the other
hand, \cite{woodmathis04} find that the emission-line spectrum of
the WIM is consistent with the harder spectral energy distributions
(SEDs) expected from density-bounded \hii\ regions, and
studies of radiative transfer in the global ISM suggest that ionizing
radiation travels over long path lengths, on the order of hundreds of pc
in the galactic plane, and 1 -- 2 kpc outside the plane 
\citep[e.g.][]{collinsrand01,zuritaetal02,seon09}.  It is also
well-known that the WIM surface brightness is highest around
\hii\ regions \citep{fergusonetal96}.

We recently developed the technique of ionization-parameter mapping
(IPM) to more directly evaluate nebular optical depth in the Lyman
continuum \citep{p12}.  This technique
uses emission-line ratio maps to determine the nebular ionization
structure, and hence, infer the optical depth.  For conventional,
optically thick Str\"omgren spheres, there is a transition zone
between the central, highly excited region and the neutral environment.
These transition zones are characterized by a strong decrease in the ionization
state, and hence, the gas ionization parameter, which is the ratio of
radiation energy density to gas density.  
Objects that are optically thick to ionizing photons reflect
stratified ionization structure, showing low-ionization envelopes 
around highly ionized central regions.  In contrast, optically thin nebulae
will exhibit weak or nonexistent lower-ionization transition zones,
and thus they show high ionization projected across the entire
object.  These usually show irregular and disrupted morphology, which is
consistent with radiation-MHD simulations by \cite{arthuretal11} for
highly ionized \hii\ regions.   

This simple IPM technique allowed us to estimate
the optical depths of the \hii\ regions in the Magellanic Clouds
using \Ha, [\oiii] $\lambda\lambda$4959,5007, and [\sii]
$\lambda\lambda$6717,6732 data from the Magellanic Clouds
Emission-Line Survey \citep[MCELS;][]{smithetal05}.
We were thus able to 
determine that optically thick nebulae dominate at low \Ha\ luminosity,
while high-luminosity objects are mostly optically thin, dominating
at luminosities above $10^{37}\ \rm erg\ s^{-1}$ in both galaxies
\citep{p12}.
This implies that most of the bright \hii\ regions observed in
star-forming galaxies are optically thin.  Similarly, we found that
the frequency of optically thick \hii\ regions strongly correlates
with \hi\ column, although at the lowest \nhi, the optically thin
objects dominate.  Thus, despite strongly differing properties of the neutral ISM of
these galaxies, the quantitative properties of the nebular radiative
transfer are remarkably similar.  Our results demonstrate that
IPM is a vivid and powerful tool for constraining the optical depth to
ionizing radiation \citep{p12}.  However, we need to further evaluate this
technique and understand it in the context of other ISM properties and
diagnostics. 

In particular, dust properties are a significant factor in the
radiative transfer of ionizing radiation, and they also offer
multifaceted probes of this process.  Polycyclic aromatic hydrocarbon
(PAH) emission is sensitive to Lyman continuum radiation and is 
destroyed by it \citep[e.g.,][]{tielens08}, while larger dust grains absorb and re-emit this
radiation.  We therefore use 8~\mic\ and 24~\mic\ data from the
{\sl Spitzer} survey of the Magellanic Clouds, SAGE [Surveying the
Agents of Galaxy Evolution; \citep{meixneretal06}], and dust maps from
\citet{gordonetal14} based on the analogous far-infrared
{\sl Herschel} survey, HERITAGE [{\sl Herschel} Inventory of The
Agents of Galaxy Evolution; \citep{meixneretal13}] to examine the
Lyman continuum radiative transfer.

\section{8 \mic\ PAH Emission}

The 8 \mic\ bandpass probes the bright, 7.7
\mic\ and 8.6 \mic\ PAH features, particularly ionized PAHs
\citep[e.g.,][]{lidraine01a}.  \citet[][2009]{bauschlicher08} 
attribute the 7.7 \mic\ band to
C-C stretch and C-H in-plane bending vibrations in small and  
large charged PAHs, and the 8.6 \mic\ emission to C-H
in-plane bending vibrations in large, charged, compact PAH molecules
($>70$ C atoms).  In the Large
Magellanic Cloud (LMC), PAH emission is typically an order of
magnitude brighter than other contributions to this band in both star-forming
and diffuse ISM \citep{bernard08}.  Even in the low-metallicity SMC,
spectral analysis of objects with low PAH fractions shows that
these emission features still dominate the continuum
\citep{sandstrom10}.

\begin{figure*}
\vspace*{-1.0in}
\epsscale{0.5}
\plotone{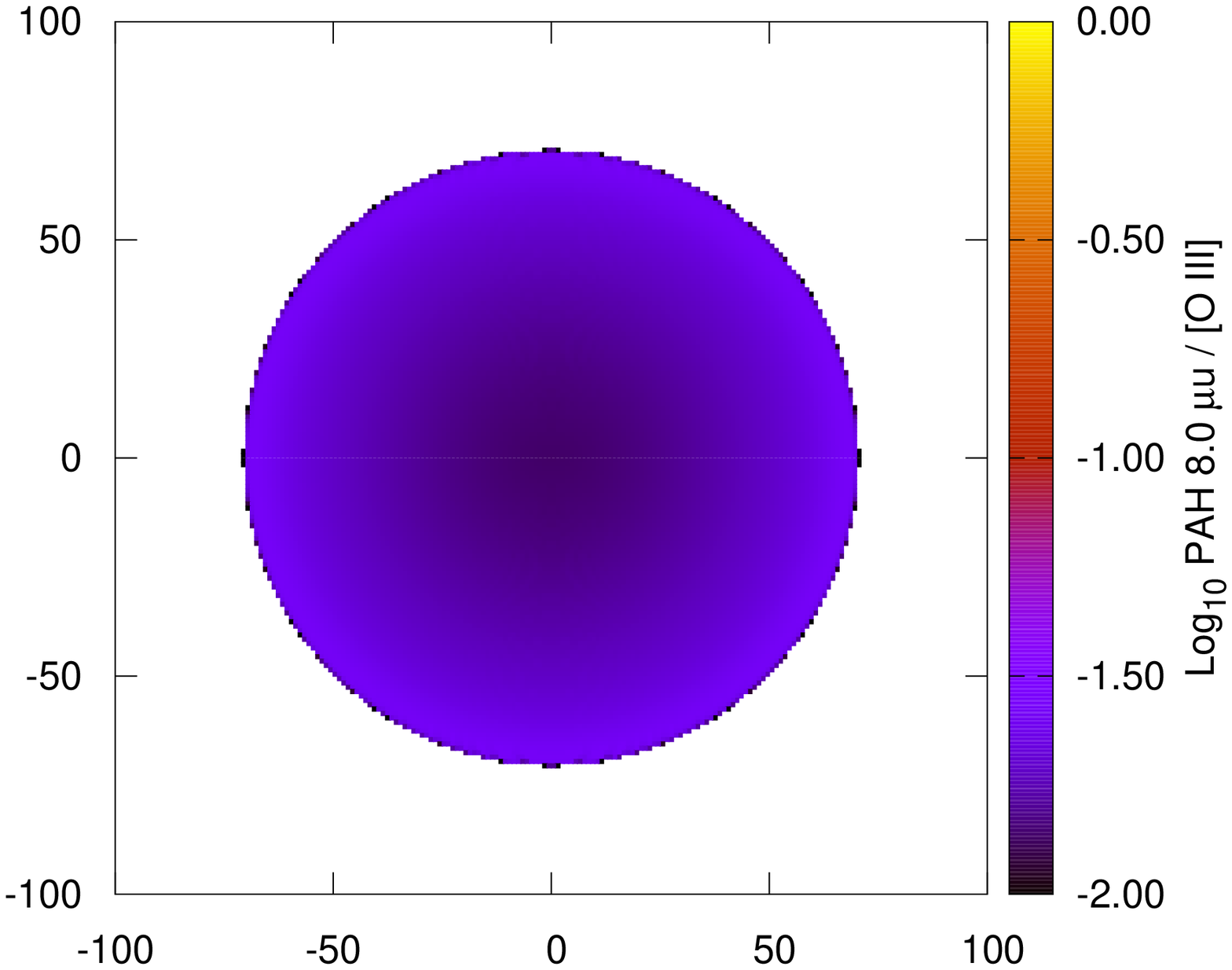}
\plotone{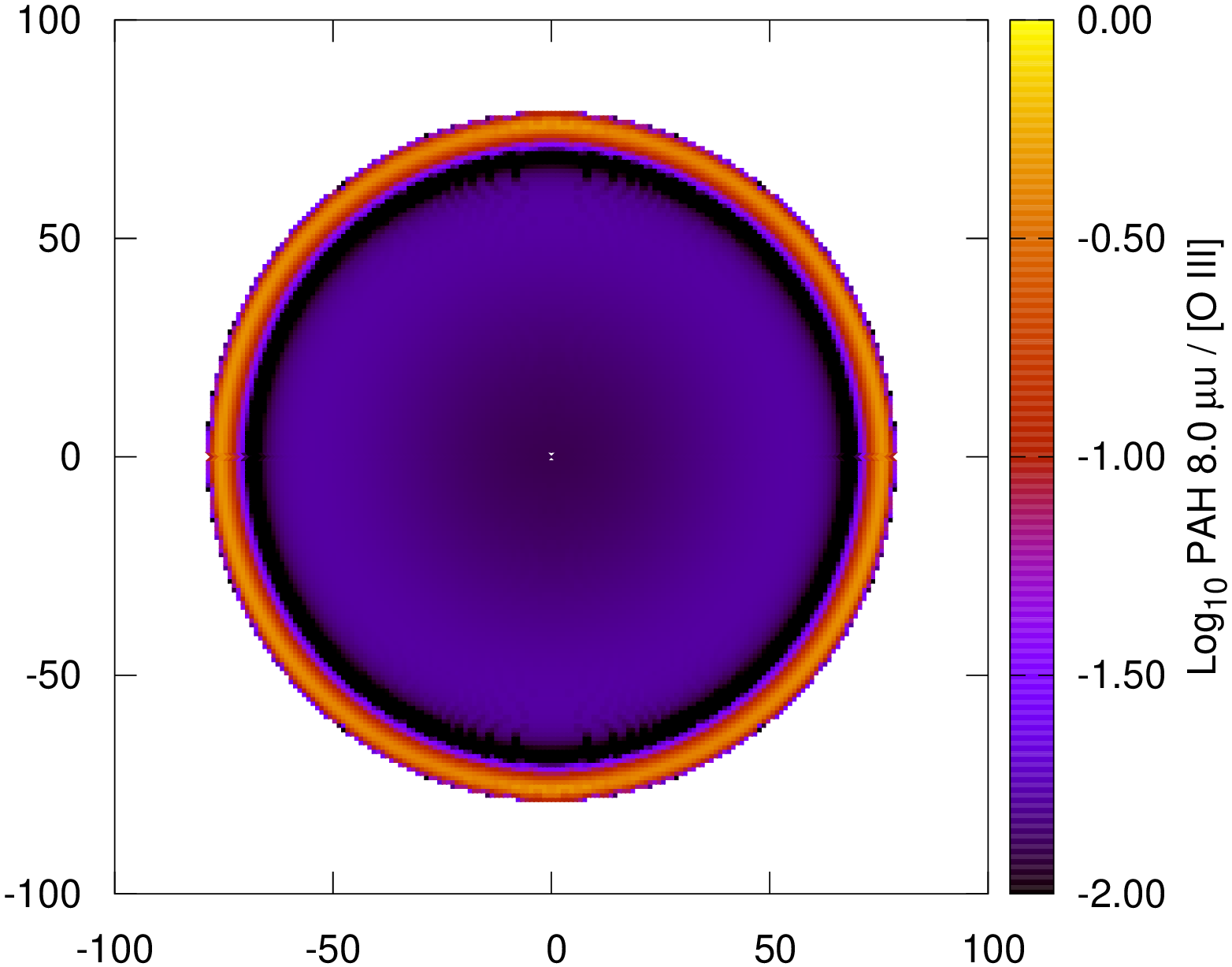}
\caption{Modeled 8\mic/[\oiii] ratio map of an LMC \hii\ region ionized by
  an O6 V star, for $\tau=0.5$ (left) and $\tau=20$ (right).  The
  the assumed parameters are the same as for Figure~2 of \citep{p12},
  with $x$ and $y$-axes showing spatial projection in arcsec at the
  LMC distance.  PAHs survive and dominate emission near the
  Str\"omgren edge in the optically thick object, in contrast to
  optically thin object.
}
\label{f_pelleg}
\end{figure*}

PAHs are generally found to be anticorrelated with ionized gas, 
indicating that they are destroyed by ionizing radiation 
\citep[e.g.,][]{povichetal07,pavlyuchenkovetal13}.  Indeed, aromatics
are a major component of the Lyman continuum opacity \citep{lidraine01}.
We therefore expect that
optically thin \hii\ regions should show less PAH emission in their
peripheries relative to optically thick objects.  Thus,
the spatial distribution of PAHs near optically thin \hii\ regions
might behave similarly to that of low-ionization atomic species.  Therefore,
mapping of 8 \mic\ PAH emission relative to a high-ionization atomic
species (e.g., [\oiii]) might yield results similar to
ionization-parameter mapping based on a low-to-high ionization ratio
map as done by \cite{p12}.  Figure~\ref{f_pelleg} shows example
8\mic/[\oiii] ratio maps of an \hii\ region simulated with {\sc Cloudy}
\citep{ferland13}.  We show an object
ionized by an O6~V star for Lyman continuum optical depths of $\tau =
0.5$ and 20.  This figure is analogous to Figure~2 of \citet{p12}, and
illustrates that, in principle, 8\mic/[\oiii] should behave
similarly to [\sii]/[\oiii].
In what follows, we use the high-quality, 8 \mic\ residual images from
the SAGE survey \citep{gordonetal11, meixneretal06}, for which the
stellar point sources were removed via PSF fitting \citep{sewilo09},
alleviating stellar contamination.

\begin{figure*}
\plotone{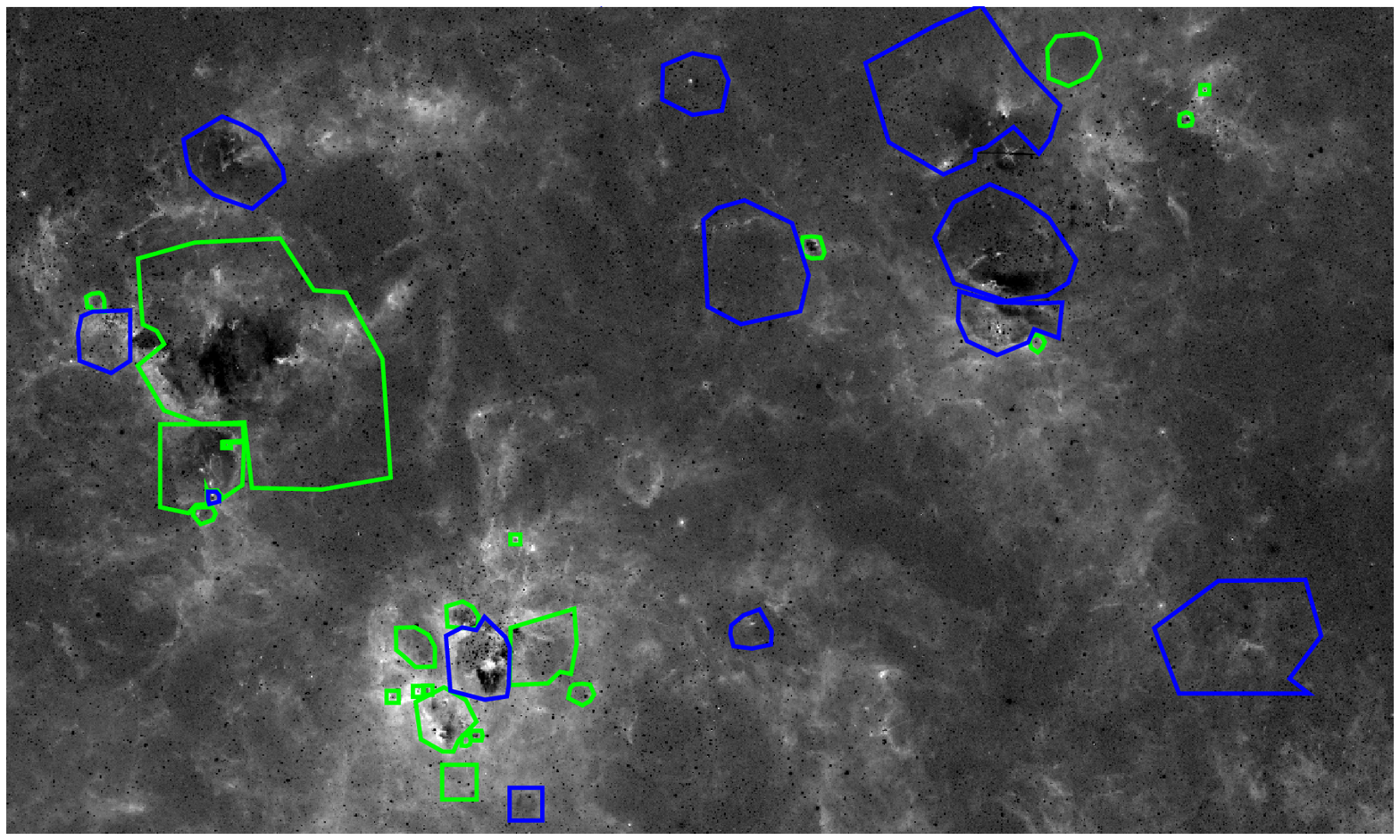}
\plotone{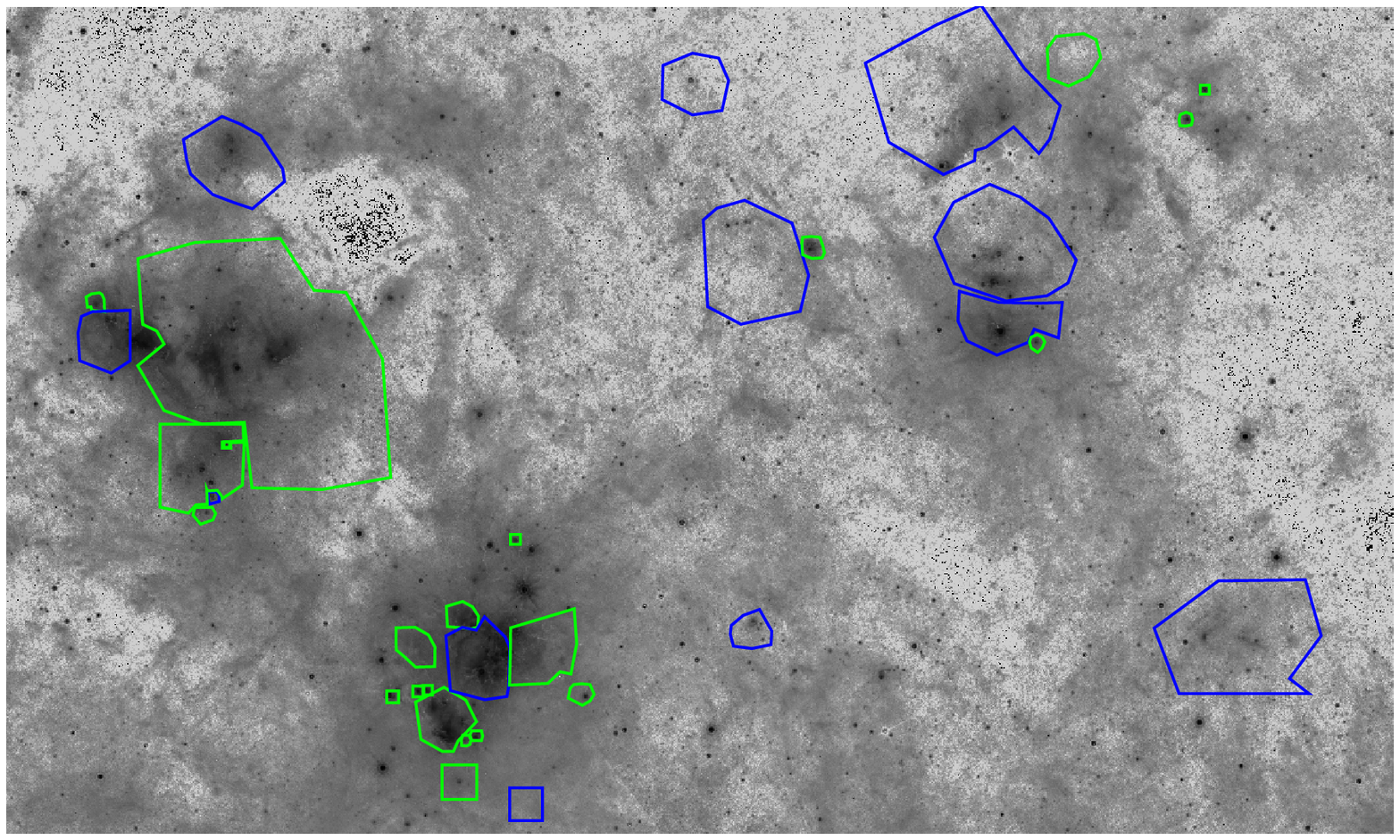}
\caption{Map of the total 8\mic/[\oiii] ratio (top) and 8\mic/24\mic\ ratio
(bottom) for a section of the LMC.  White indicates larger values.
The polygons show the nebular boundaries from \citet{p12}.  Objects
classified as optically thick and thin in that work are shown with
green and blue polygons, respectively. 
}
\label{fig_8m2o3ratio}
\end{figure*}

Figure~\ref{fig_8m2o3ratio} (top panel) shows the 8$\mu$m/[\oiii]
ratio map for a region in the LMC, constructed from the continuum-subtracted
SAGE image and the [\oiii] image from the MCELS survey \citep{smithetal05};
white indicates high values.  The apertures defining the \hii\ regions
from \cite{p12} are 
overplotted, with green and blue showing optically thick and thin
objects, respectively, as determined by IPM in that work. 
Figure~\ref{fig_8m2o3ratio} shows that objects previously identified as
optically thin tend to show less PAH emission compared to those
identified as optically thick.  

Using the same continuum-subtracted images, we measured the 8
\mic\ flux densities of the \hii\ regions using
Funtools\footnote{http://hea-www.harvard.edu/RD/funtools/} 
routines for ds9.  This was done for all the objects catalogued 
as optically thick or thin, including ``blister'' regions, by
\cite{p12}, using the apertures defined in that work.
These apertures are defined based on the nebular emission and
ionization structure, and we note that physically associated 8
\mic\ flux may not always correlate well with
the aperture boundaries.  We tried to determine a systematic method to
modify the apertures to avoid this problem.  However, the 8
\mic\ spatial morphology varies strongly from that of the nebular
emission and is fraught with confusion from background and
neighboring emission.  Thus, there is no obvious way to redefine the
apertures to accurately define the boundaries between physically
associated and unassociated emission for most objects.  We caution that
the 8 \mic\ flux density measurements across the samples are therefore subject
to larger uncertainties in terms of their association with the
specified \hii\ regions.  It is hard to quantify these uncertainties,
but they can be on the order of 50\% for some objects, and much less
for others.  

\begin{figure}
\plotone{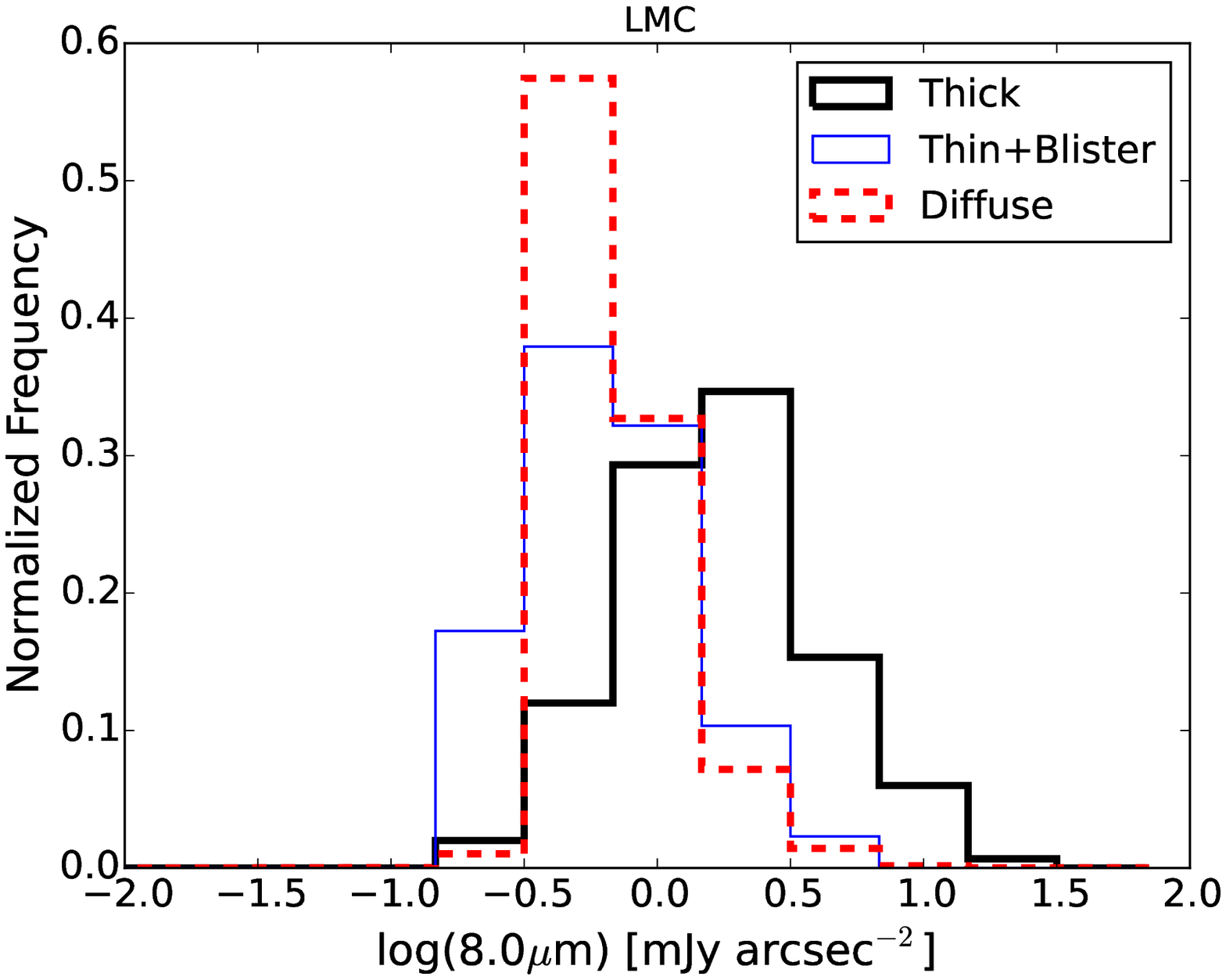}
\plotone{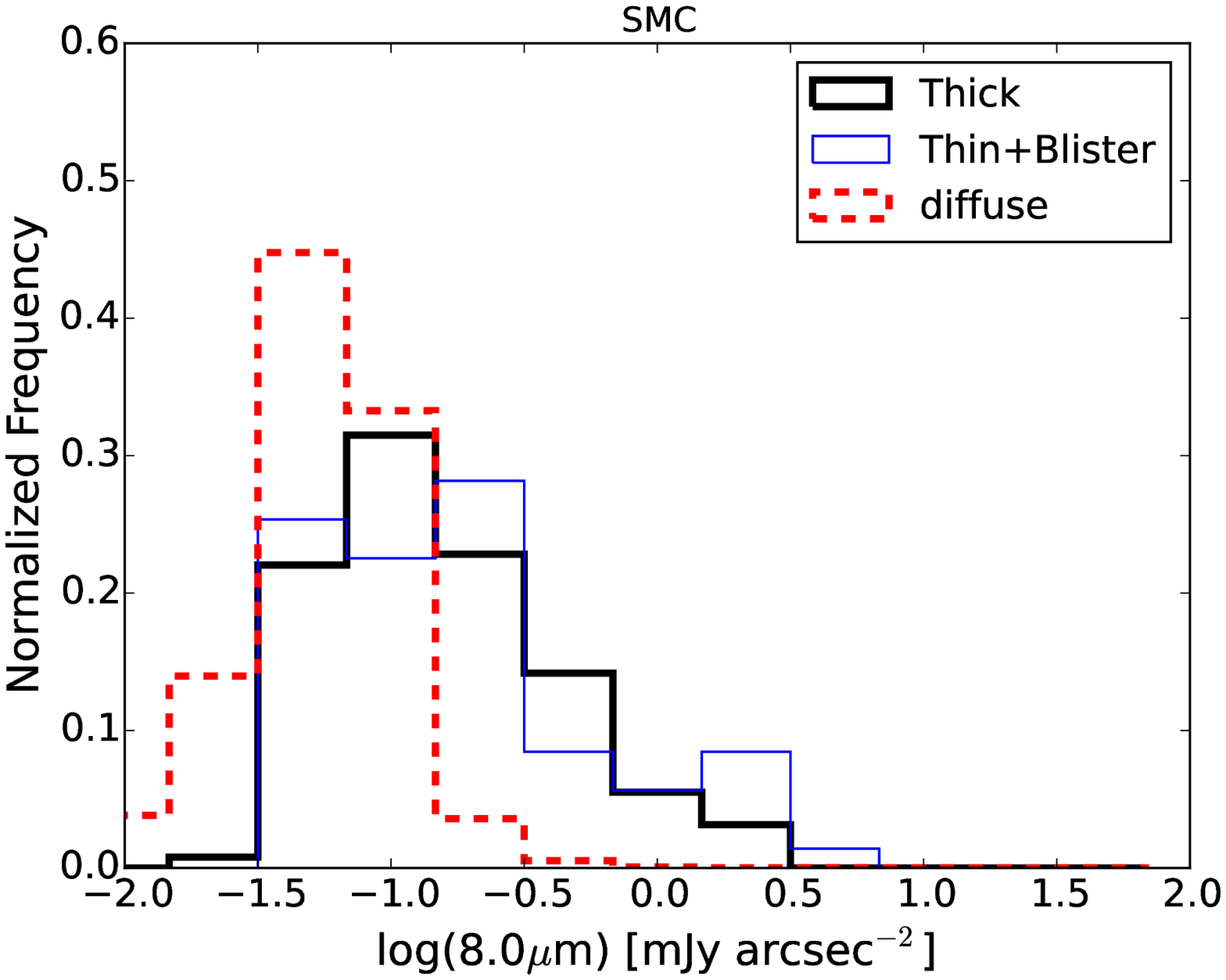}
\caption{The 8.0 \mic\ surface brightness distributions for optically thick 
(black thick line) and thin (blue thin line) \hii\ regions.  The
distribution for the diffuse, background emission
(red dashed line) is also included.  The top
and bottom panels show the LMC and SMC, respectively.}
\label{fig_my80mdists}
\end{figure}

\begin{figure*}
\plotone{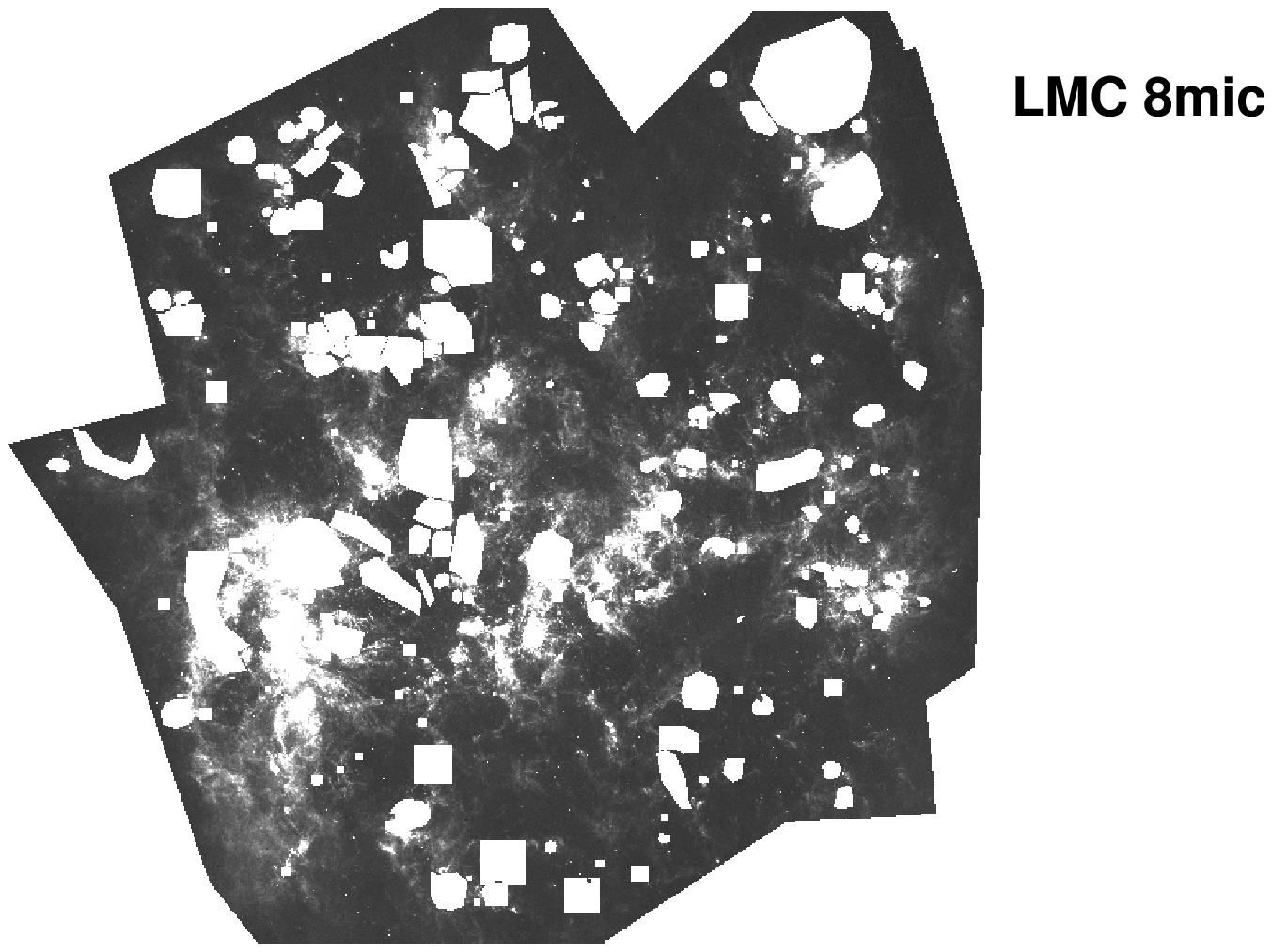}
\plotone{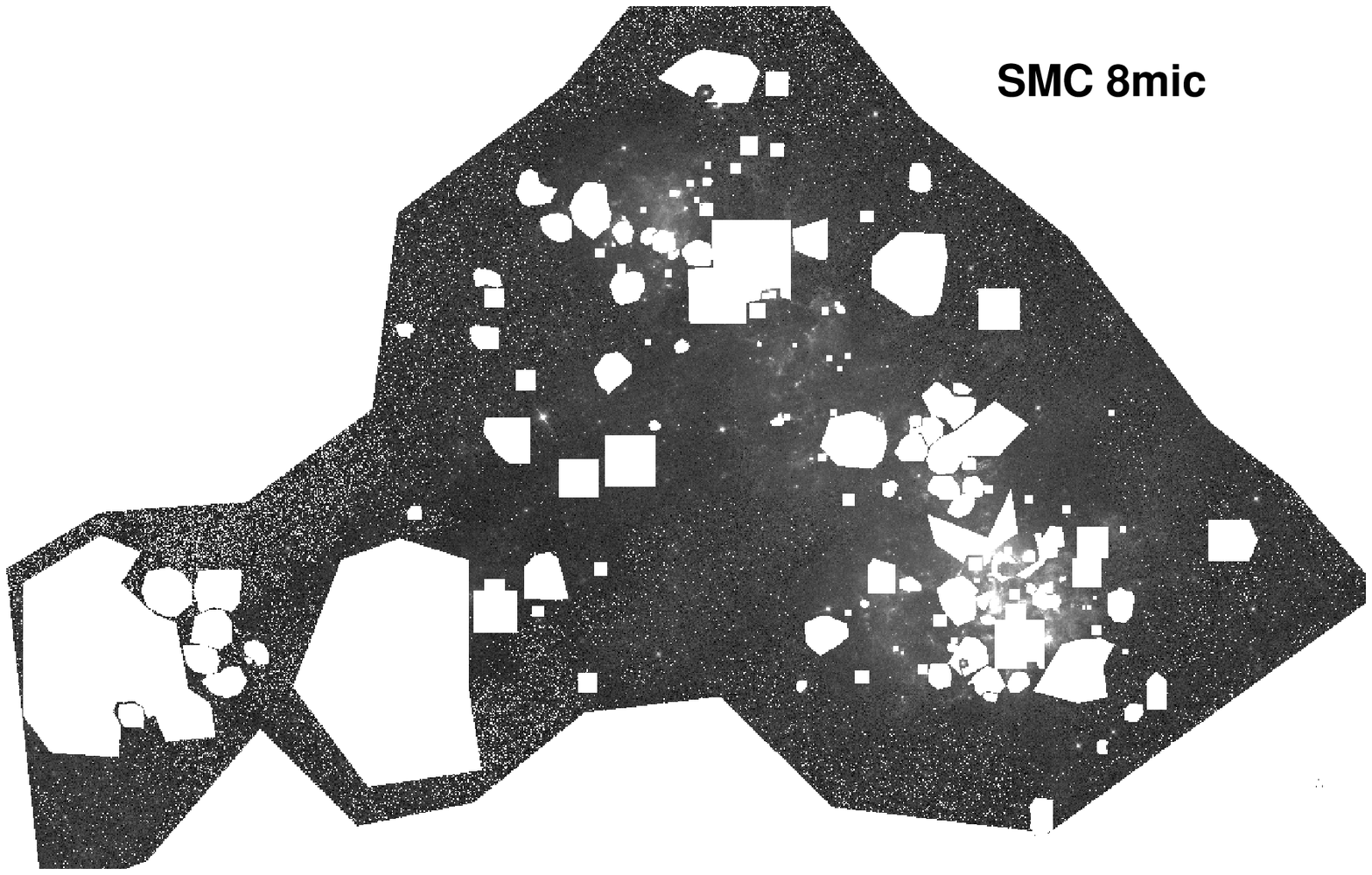}
\caption{Maps of the 8 \mic\ emission with the \hii\ regions masked, for determining the 
background emission.}
\label{fig_masks8}
\end{figure*}

Figure~\ref{fig_my80mdists} shows the 8 \mic\ flux surface
brightness distributions for the \hii\ regions in the LMC (metallicity
$0.6Z_\odot$) and SMC ($0.25Z_\odot$; Russell \& Dopita 1992),
respectively.  Objects identified as optically thick by \cite{p12}
are shown with thick lines, and those identified as
optically thin by thin lines.  Figure~\ref{fig_my80mdists} also
shows the distribution of background, diffuse 8~\mic\ emission (dashed
lines) for each galaxy, defined by the regions shown in
Figure~\ref{fig_masks8}.  
It is apparent in the upper panel of Figure~\ref{fig_my80mdists}
that the candidate optically thick objects show more
8 \mic\ emission than candidate optically thin ones in the LMC, which
is consistent with the destruction of PAHs by the Lyman continuum
radiation.  This is also confirmed by the fact that the optically thin
objects are seen to be at the background levels.

In contrast, the lower panel of Figure~\ref{fig_my80mdists} shows that in the
SMC, the 8 \mic\ surface brightness distributions for the optically 
thick and thin objects are essentially the same:  for
the optically thick objects, the median 8 \mic\ surface brightness in
the LMC is 1.2 mJy arcsec$^{-2}$, while in the SMC, it is much
lower, only 0.18 mJy arcsec$^{-2}$.  This is likely linked to 
the extremely low PAH emission found in low-metallicity
environments \citep[e.g.,][]{madden06,wu06,engelbracht05}, which is due
to actual low PAH abundance in these conditions
\citep{draine07,munozmateos09}.  \citet{sandstrom10} examined the
spatially resolved PAH abundance across the SMC, confirming the overall
low PAH fraction, but finding strong differentiation between
molecular clouds and diffuse ISM, with clouds showing PAH
fractions 2 -- 3 times higher than diffuse gas.  This resolved study
points to a model in which these aromatics form within molecular
clouds via photoprocessing in the mantles of larger dust grains 
\citep{greenberg00}; the PAHs are subsequently destroyed by
stellar UV radiation, which is less inhibited by dust in low
metallicity environments \citep[e.g.,][]{madden06,gordon08}.
PAH destruction is further enhanced by their smaller average sizes, as
found in the SMC by \citet{sandstrom12}.  This contrasts with PAH
abundance models at higher metallicity in which additional processes
contribute to PAH production, and dustier environments inhibit the
propagation of UV radiation \citep[e.g.,][]{paradis09}.  The large
observed variation in PAH abundances of star-forming regions in the
SMC is thus modulated by their remaining molecular gas, and the local UV
photon flux or ionization parameter.  This model is consistent with
the observed presence and variation of the 2175 \AA\ bump in the SMC
B1-1 cloud \citep{maizapellaniz12}.  If the PAH production
indeed depends on the existence of larger dust grains, it is 
necessarily much lower in metal-poor environments.  Thus, our results
in Figure~\ref{fig_my80mdists} can be understood such that the
large stochastic variation in PAH abundance masks any
systematic differences between optically thick and thin \hii\ regions.

\begin{figure*}
\plotone{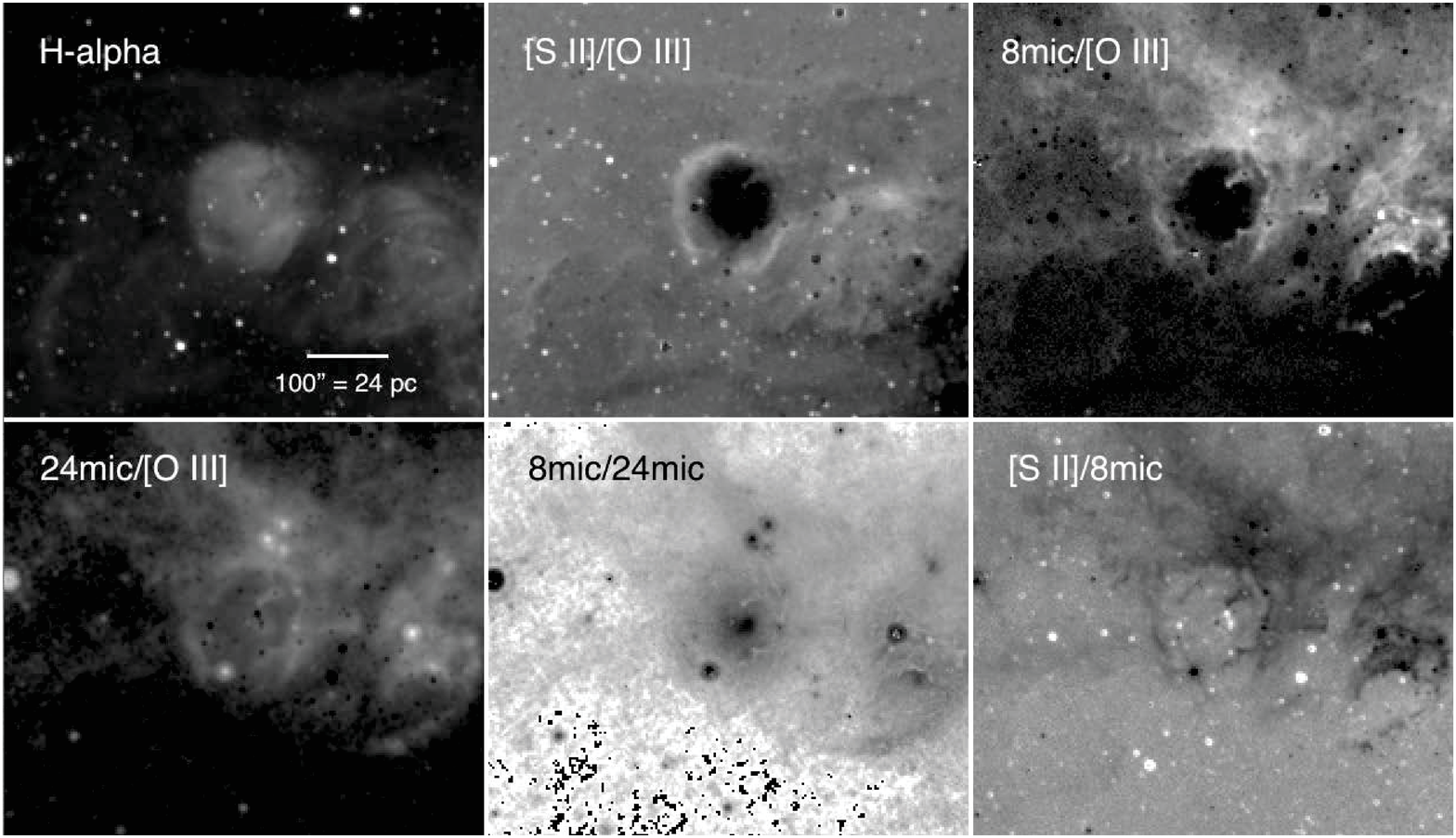}
\caption{MCELS-L215, an optically thick region in the LMC.  The panels
  show the qualitative morphology in
  H$\alpha$, [\sii]/[\oiii], 8\mic/[\oiii], 24\mic/[\oiii],
  8\mic/24\mic, and [\sii]/8\mic\ with white showing higher values on
  a logarithmic scale.} 
\label{fig_L215}
\end{figure*}

\begin{figure*}
\plotone{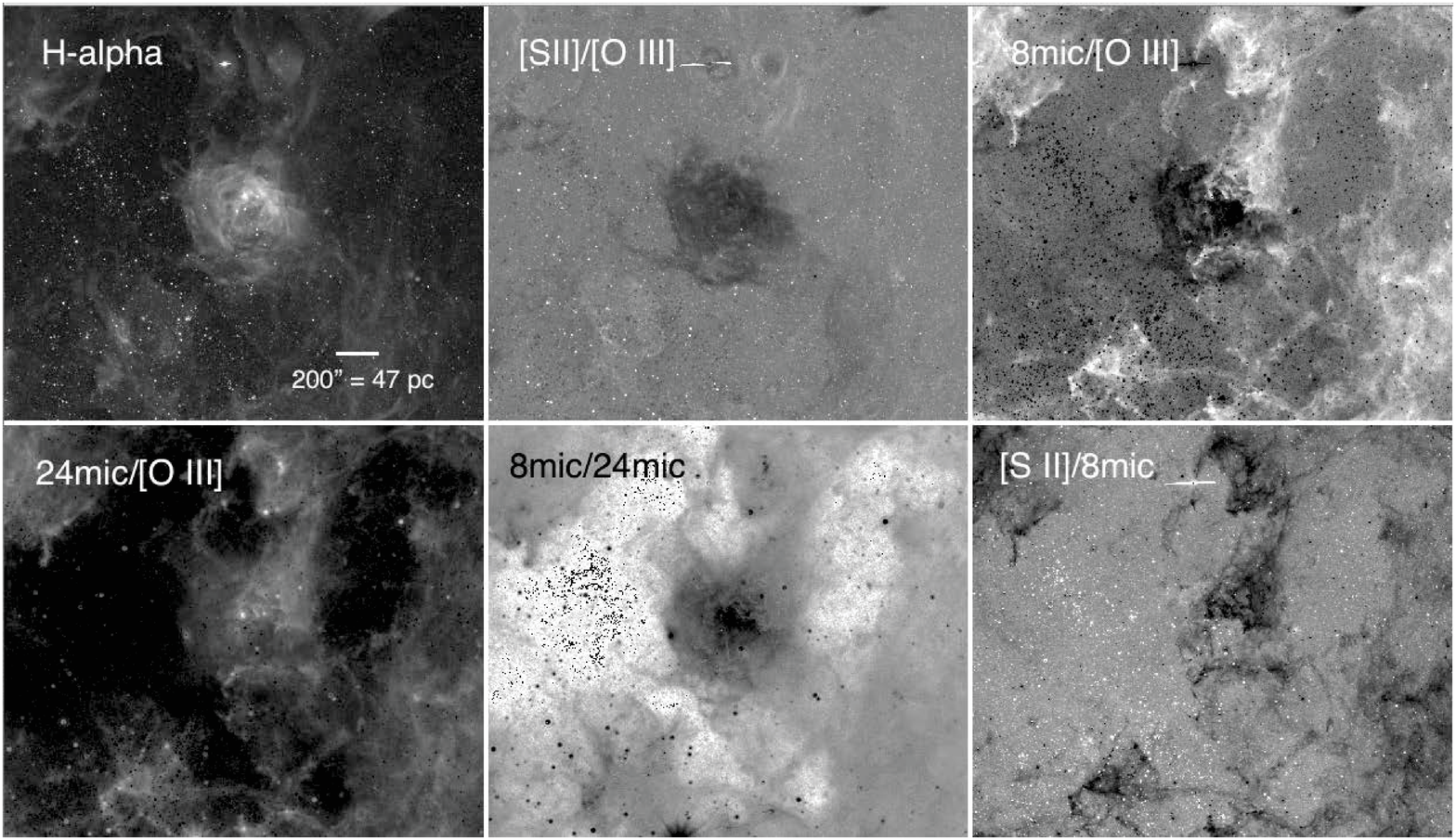}
\caption{MCELS-L258, an optically thin region in the LMC, shown with the
  same imaging as in Figure~\ref{fig_L215}. }
\label{fig_L258}
\end{figure*}

Can 8 \mic\ PAH imaging be useful for estimating the nebular optical
depth when combined with, for example, mapping in a high-ionization
atomic species?  This would be similar to the IPM technique based on
[\sii]/[\oiii] mapping.  For objects with at least LMC metallicity, the data
suggest that the 8 \mic\ imaging can provide valuable information.  At lower metallicity, as
seen in the SMC, PAHs are not abundant enough to be used
for such a diagnostic.  A couple of example objects from the LMC are
shown in Figures~\ref{fig_L215} and \ref{fig_L258}, which show 
region MCELS-L372 (optically thick) and MCELS-L258 (optically thin) in
\Ha, [\sii]/[\oiii], 8\mic/[\oiii], 24\mic/[\oiii], and 8\mic/24\mic.
There is similarity between the [\sii]/[\oiii] and
8\mic/[\oiii] ratio maps, although we also see that the
8 \mic\ emission extends beyond the nebular boundaries defined for the
regions.  In many cases it also appears morphologically unrelated to
the \hii\ region, as in MCELS-L258 (Figure~\ref{fig_L258}).  We can
therefore expect that evaluating the optical depth based only on
8\mic/[\oiii] will not be as straightforward as when using only nebular
atomic lines.

We reclassified all the LMC objects by visual inspection 
of the regions, following the \citet{p12} methodology, but using the 8\mic/[\oiii] 
map instead of [\sii]/[\oiii], and allowing consideration of PAH
emission outside the nebular boundaries specified by \citet{p12}.
We also imposed a threshold value of 0.5 and 0.3 in these ratio maps for
the LMC and SMC, respectively, above which the objects are
considered optically thick.  Our classifications are listed in the Appendix.
We then compare with the objects' classifications by \citet{p12} as optically
thick or thin (including blister) based on [\sii]/[\oiii] maps.
The sample for which this comparison can be done corresponds to 
almost two-thirds of the objects (256 out of 401 objects) in the 
LMC, since a number of objects were either not classified as
optically thick or thin by \cite{p12} or by us, or did not correspond to
adequate detection in 8 \mic.  We find that of the 256 objects, 185
(72\%) maintain the same classifications and 71 objects (28\%)
switch classification from optically thick to thin (59 objects) or
vice versa (12 objects).

In the SMC, however, more objects change their classification (115 objects
out of 189, or 61\%) than remain the same (74 objects, or 39\%) when
evaluated with PAH emission.  This again suggests that 
PAHs are simply not abundant enough in this galaxy to provide
a useful diagnostic of radiative transfer.  However, in the LMC, for
objects whose classifications are consistent for both nebular and
PAH-based methods, the 8 \mic\ data can provide important confirmation.

As in the LMC, we also find in the SMC that more objects switch
classification from optically thick to thin (88) than vice versa
(27).  This trend is consistent with PAHs being a more sensitive 
indicator of UV flux than low-ionization atomic species.  As discussed
by \citet{p12}, although it usually indicates optically thick
conditions, the presence of a low-ionization envelope is also seen in
some optically thin objects, especially those with softer ionizing
sources.  The nebular-based classifications therefore might
discriminate at somewhat higher optical depths than the PAH-based ones.
More data is needed to determine how much of the discrepancy
between the methods is due to this effect, and
how much is due to errors caused by PAH spatial distribution,
background confusion, and lower spatial resolution in the
8 \mic\ image, as well as misclassifications from the nebular lines.

\section{24 \mic\ Hot Dust emission}

Very small dust grains within \hii\ regions absorb energetic photons produced by the 
massive stars and re-emit this energy in the 24 \mic\ band, which is 
an indicator of hot dust \citep[e.g.,][]{draineli07}.  Hence 
24 \mic\ emission has been used as a tracer of obscured star formation 
\citep[e.g.,][]{calzettietal07}.  Optically thick objects, with
higher gas-to-photon densities, might be expected to have more dust,
and thus, correspondingly stronger 24~\mic\ emission.  However, we
note that these dust grains, which are on average larger than PAHs,
are not as easily destroyed by UV radiation.  Thus, they tend to
associate with individual dense knots, and also remain somewhat more
uniformly distributed in the star-forming regions than 
PAHs.  This is seen in the spatial distribution of 24 \mic\ emission in
Figures~\ref{fig_L215}, and \ref{fig_L258}. 

\begin{figure}
\plotone{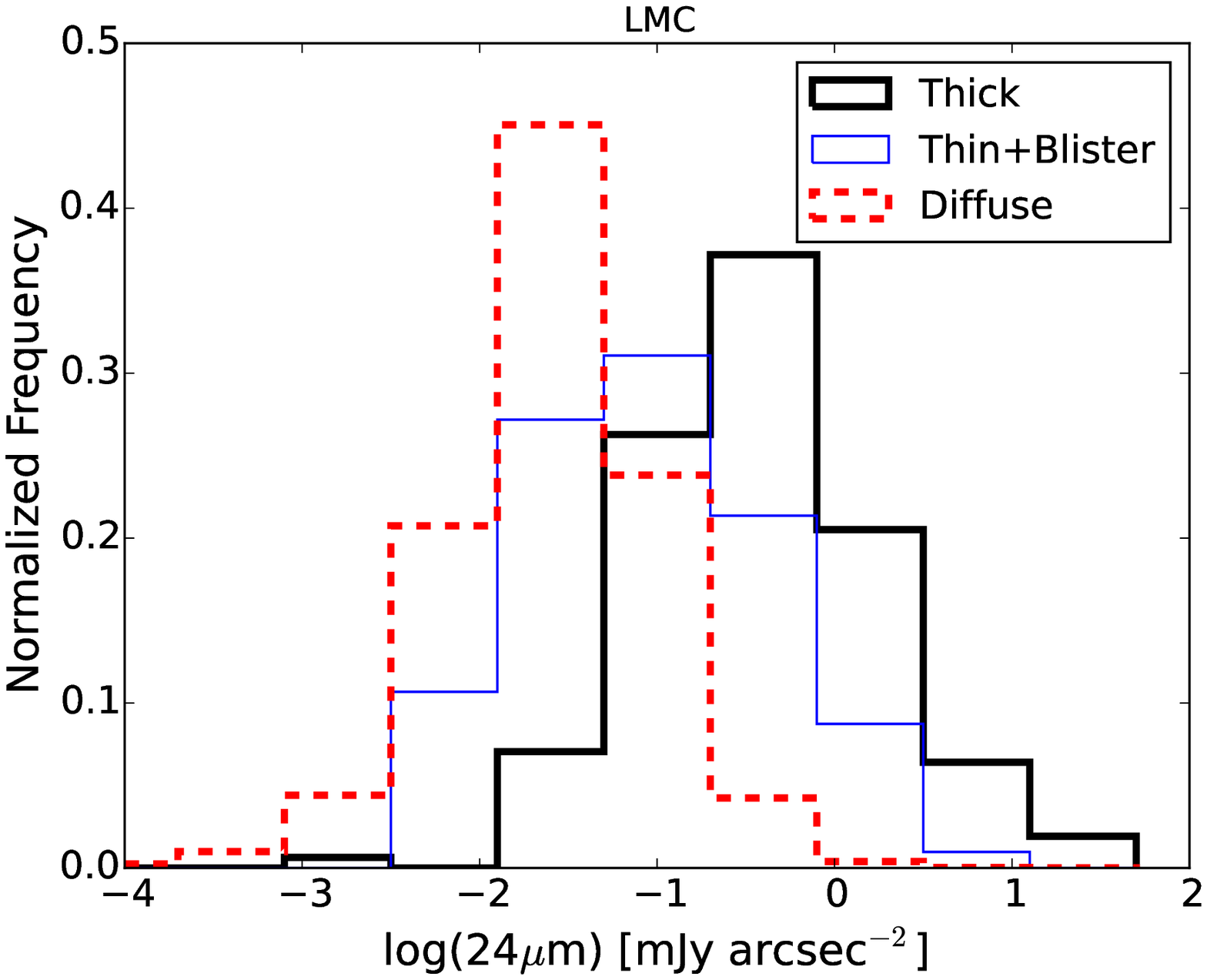}
\plotone{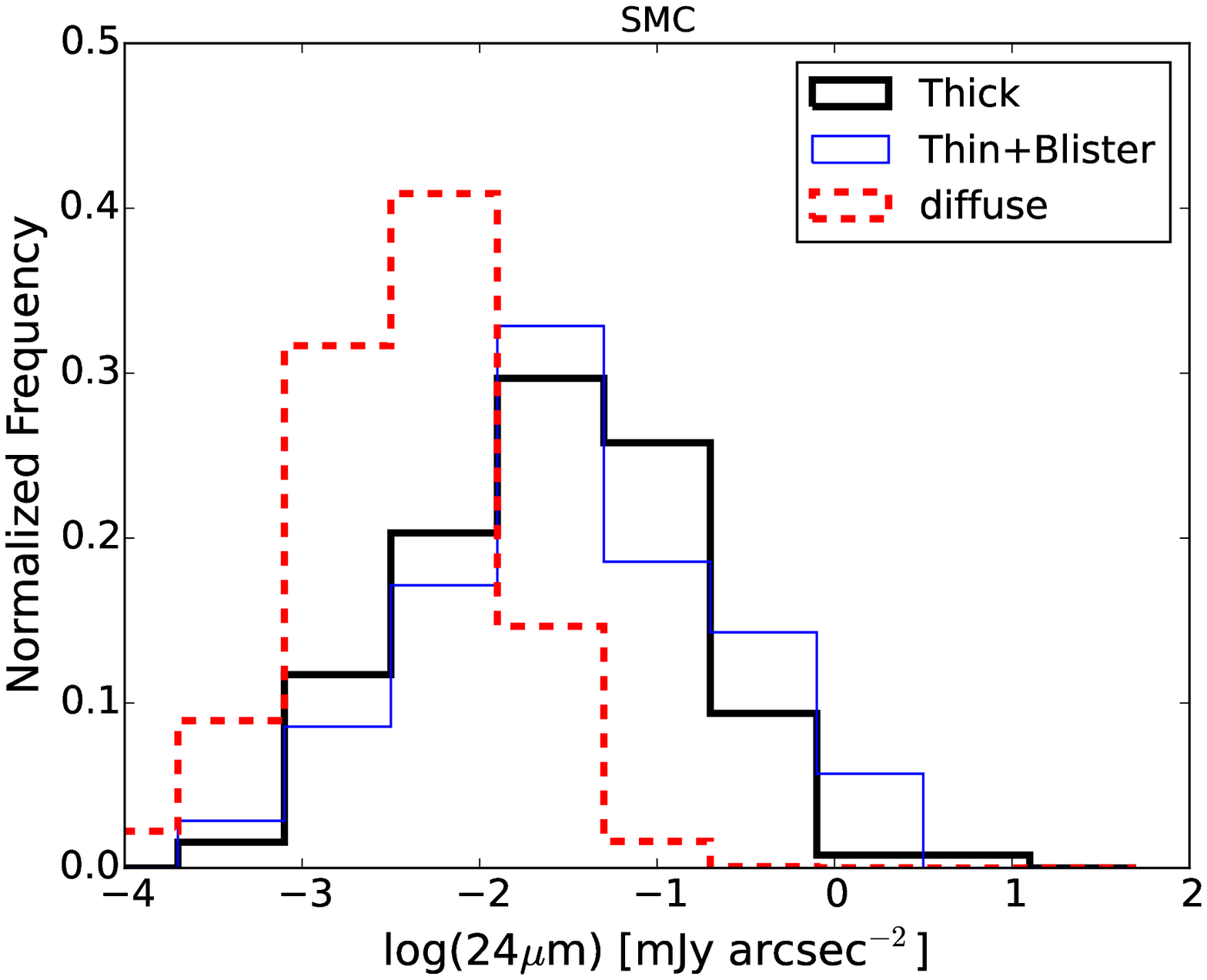}
\caption{Distribution of the 24 \mic\ surface brightness for the LMC (top) and the SMC
(bottom).  Line types are as in Figure~\ref{fig_my80mdists}.  \label{fig_my24mdist}}
\end{figure}

We measure the 24 \mic\ surface brightnesses for our sample objects in
the same way as for the 8 \mic\ emission.  The 24 \mic\ data are not
continuum-subtracted, since there is no significant stellar 
continuum contributing to this band.  Figure~\ref{fig_my24mdist}
shows the 24 \mic\ surface brightness distributions for the LMC (top)
and SMC (bottom).  We see that, as expected, optically thick
regions in the LMC have higher 24 \mic\ surface brightness than
optically thin ones.  The median values are 0.44 and 0.13 mJy
arcsec$^{-2}$ for  the thick and thin regions, respectively, in this
galaxy. 

However, for the SMC, the 24 \mic\ surface brightness distributions are
essentially the same for the optically thick and thin objects
(Figure~\ref{fig_my24mdist}).  As in the case of the 8~\mic\ 
emission, this is likely due to the low SMC metallicity and hence, low
dust content, as well as generally lower ISM density relative to the LMC.
The mean 24 \mic\ surface brightness for the thick and thin regions in the SMC is 
about 0.05 mJy arcsec$^{-2}$, an order of magnitude lower
than the values for the LMC.  We do note that the diffuse background
is still slightly lower than in the \hii\ regions.

\section{Dust Mass}

We interpret our findings above to suggest that the SMC is simply too
metal-poor to sustain enough dust, both PAHs and larger grains, to generate
differential trends between optically thick and thin \hii\ regions as seen
in a more metal-rich environment like the LMC
(Figures~\ref{fig_my80mdists} and \ref{fig_my24mdist}).  To evaluate
this possibility, we use the dust map constructed by
\citet{gordonetal14} to measure the integrated dust masses using the
same method as before.

\begin{figure}
\plotone{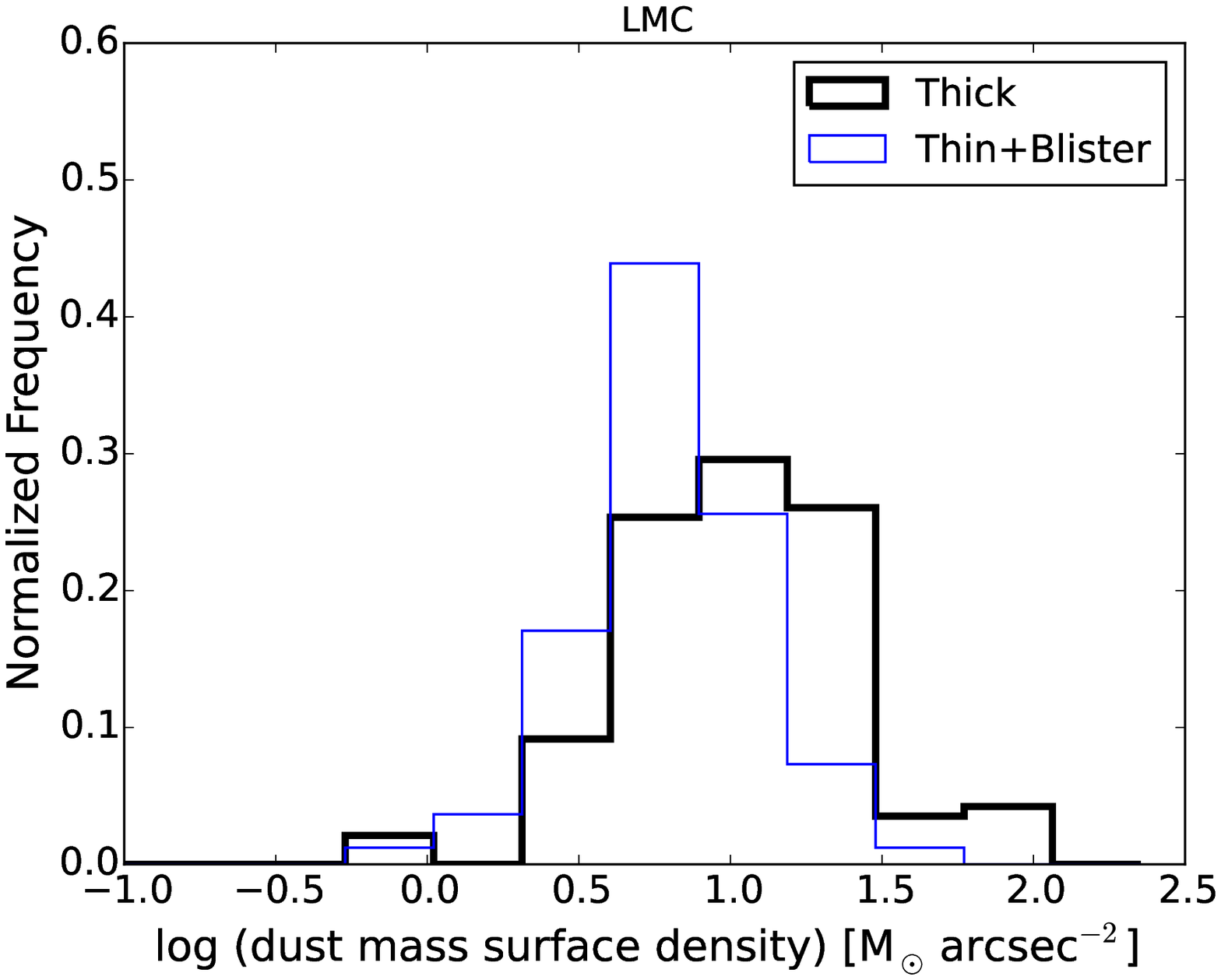}
\plotone{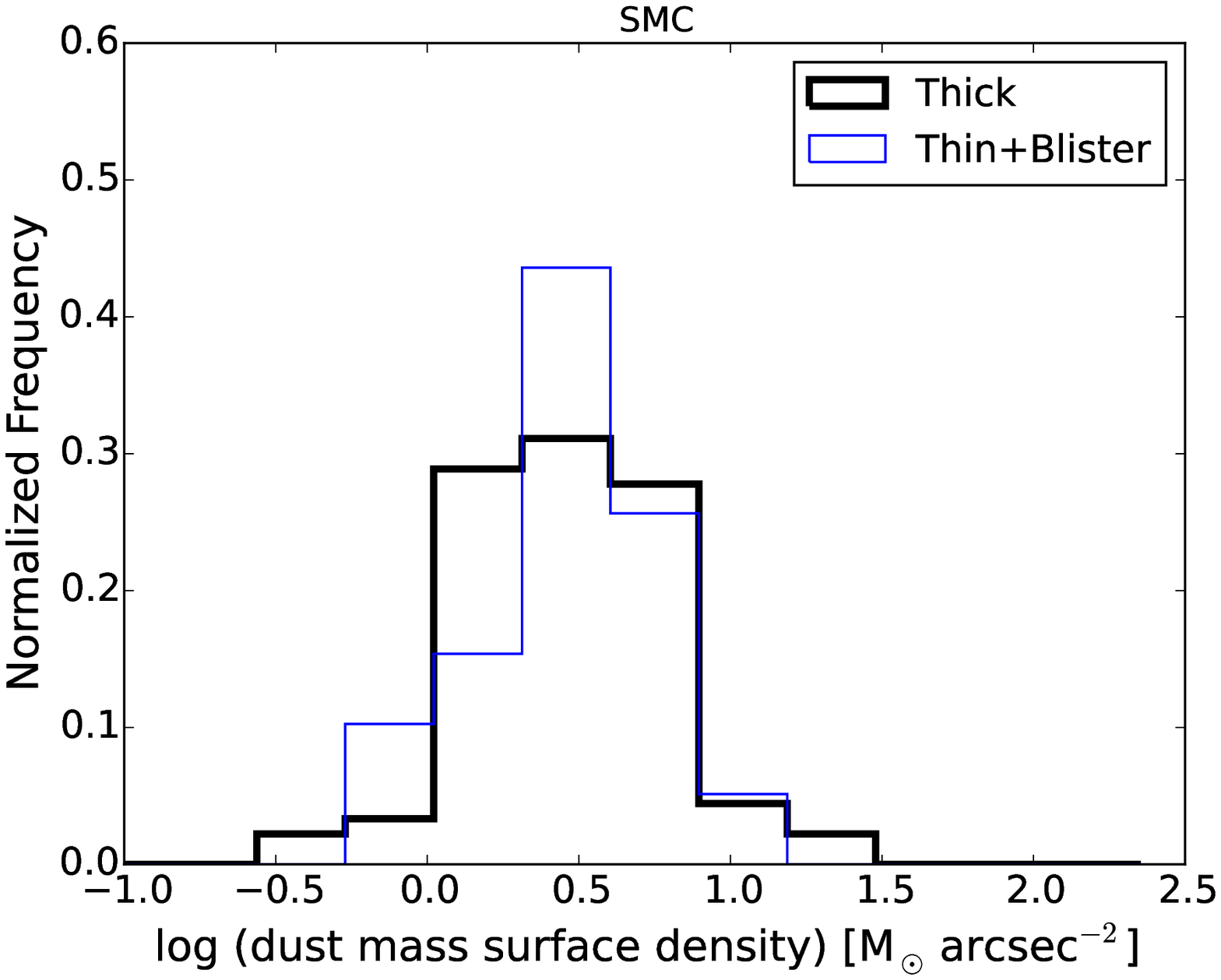}
\caption{Dust surface density distributions, with line types as before.}
\label{fig_dustmassdist}
\end{figure}

In the SMC, 129 (63\%) of 203 objects are detected, whereas in the LMC
220 (83\%) of 262 objects are detected in the dust maps. 
For objects with detections, Figure~\ref{fig_dustmassdist} shows the
distribution of dust mass surface density \Sigd\ for the optically thick and thin
objects in each galaxy, analogous to the earlier distribution plots.
The top panel of Figure~\ref{fig_dustmassdist} indeed confirms that optically thick
objects in the LMC have 1.6 times higher median \Sigd\ than their optically
thin counterparts; the median \Sigd\ are
$5.0\times10^{-3}$ and $3.0\times10^{-3}\ \rm M_\odot\ pc^{-2}$, respectively.
In contrast, there is no differentiation between optically thick and
thin objects in the SMC:  $1.1\times10^{-3}$ and $1.2\times10^{-3}\ \rm
M_\odot\ pc^{-2}$, respectively.  This value may well correspond to a diffuse
background emission, and Figure~\ref{fig_dustmassdist} may imply
that optically thin \hii\ regions have negligible \Sigd.

\begin{figure}
\plotone{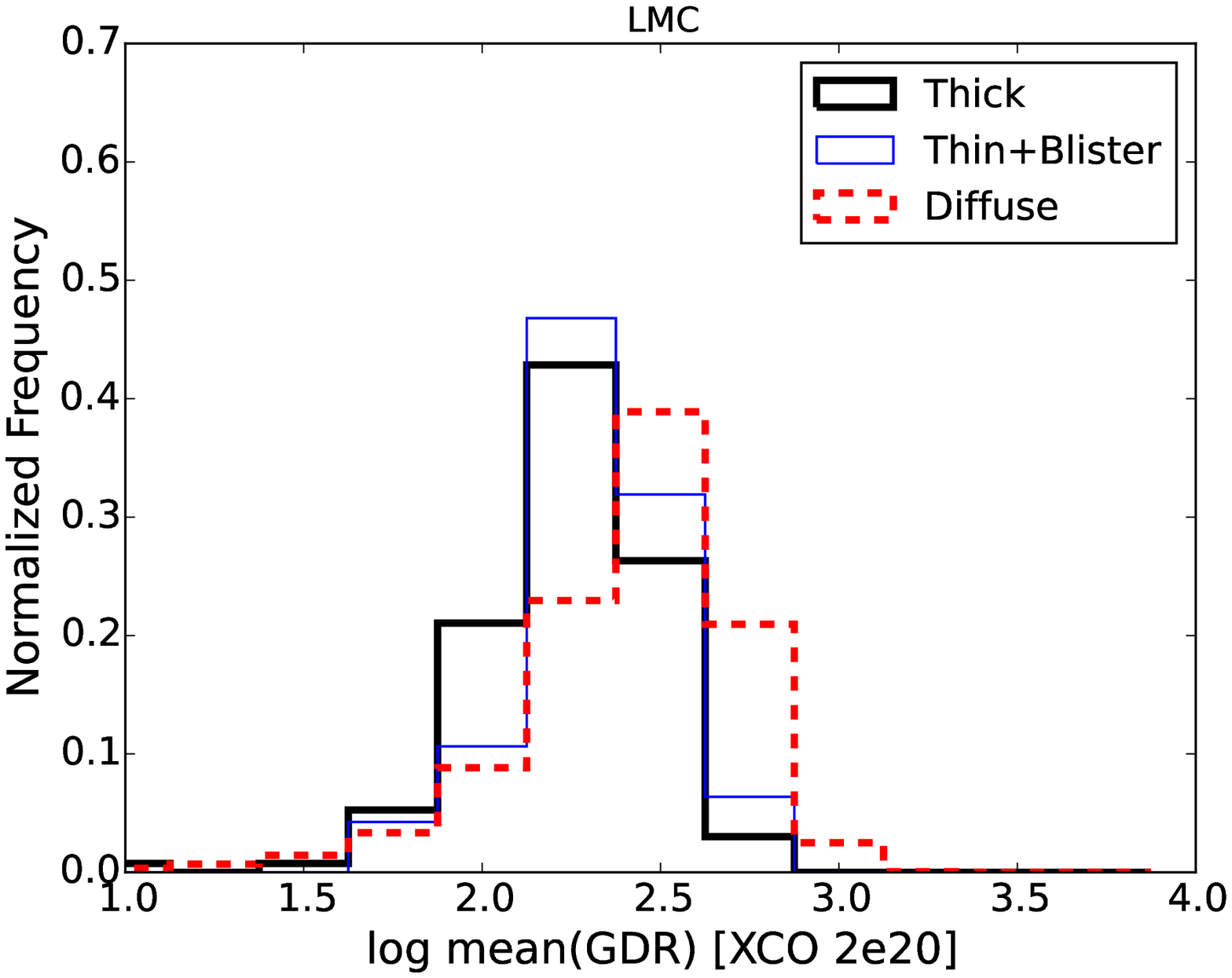}
\plotone{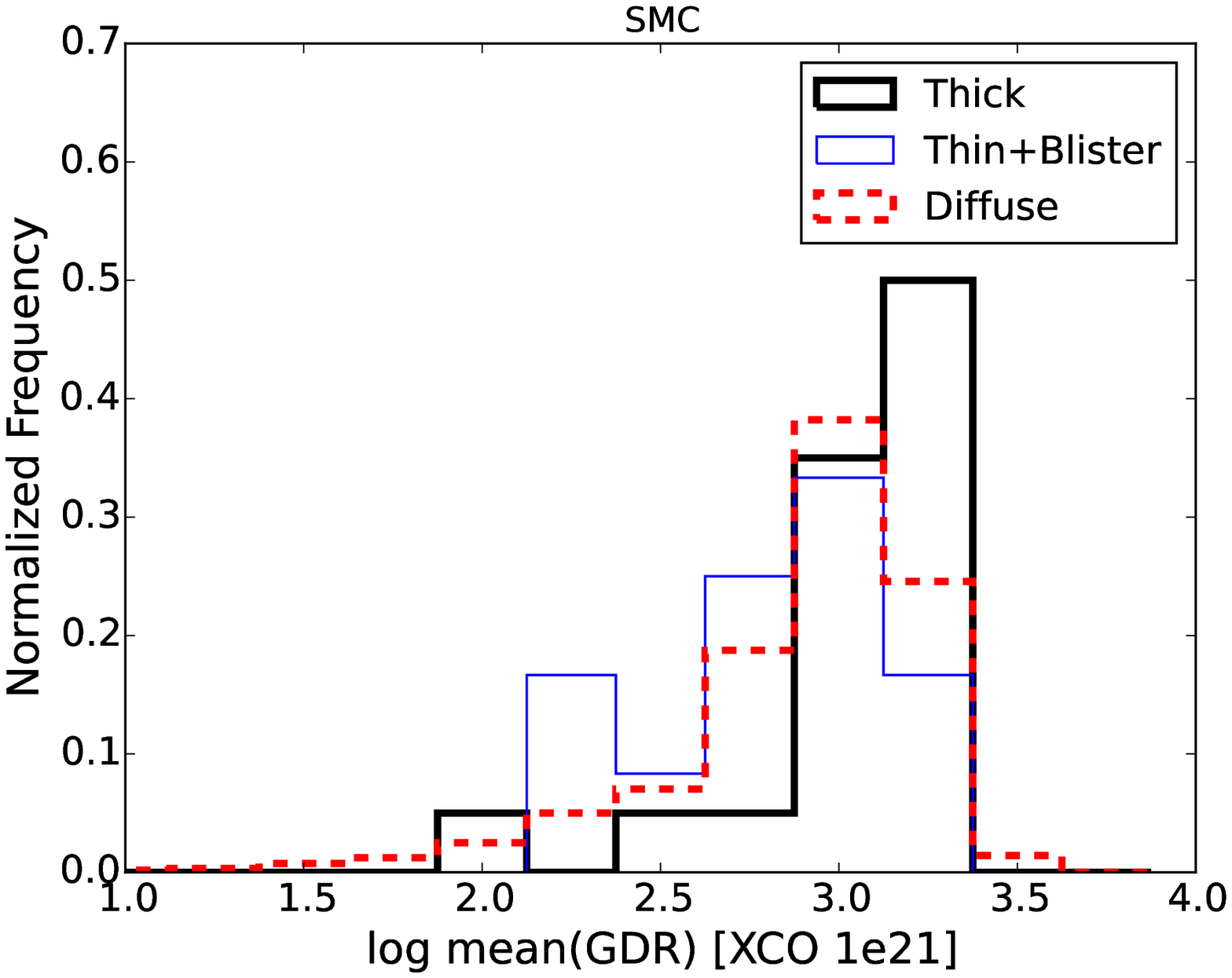}
\caption{Distributions of gas-to-dust ratios, with line types as before.}
\label{fig_gdr}
\end{figure}

These trends are further confirmed by the gas-to-dust ratios (GDR)
obtained in the same apertures.  We computed these using the GDR maps of
\citet{romanduval14}, where the dust surface density is derived from
the HERITAGE data used above \citep{gordonetal14}, and the gas surface density
includes both \hi\ \citep{kim03,stanimirovic99} and H$_2$, inferred
from CO \citep{wong11,mizuno01}.  We adopt the maps with CO-to-H$_2$
conversion factors of $X_{\rm CO,20} = 2$ and 10 in the LMC and SMC,
respectively \citep{bolatto13}.  Figure~\ref{fig_gdr} shows the
distribution in GDR for the optically thick and thin objects in both
Magellanic Clouds, analogous to our previous figures, along with the
diffuse emission.  We see that, in the LMC, optically thin objects tend to have
higher GDR than the optically thick objects, although the
effect is not dramatic.  The mean values for thin and thick objects
are 265 and 243, respectively, and the optically thin distribution is
again intermediate between the optically thick objects and diffuse gas, as seen in
Figures~\ref{fig_my80mdists} and \ref{fig_my24mdist}.  This behavior
is consistent with the conventional correlation between dust and
optically thick conditions.

As before, the behavior is different in
the SMC, now with the optically thick 
objects showing significantly higher GDRs than the optically thin
objects:  the mean values are 1701 and 959 for the two samples,
respectively.  Figure~\ref{fig_gdr} shows that the optically thin
distribution peaks at similar GDR as the diffuse gas, supporting our
premise that the dust abundance in these objects is similar to that of
the diffuse ISM, since both are governed by destruction from the UV
interstellar radiation field \citep{madden06,gordon08}.  The higher
GDR for optically thick objects may be attributable to the fact that
these \hii\ regions are also subject to UV radiation, but must
have higher gas masses necessary for optically thick \hii\ regions.

\section{Conclusion}

We have examined the 8~\mic\ PAH and 24~\mic\ hot dust emission
associated with \hii\ regions in the Magellanic Clouds to evaluate how
the emission in these bands relates to the nebular optical depth in the
Lyman continuum.  Specifically, we examined IR emission and dust
properties derived from the SAGE and HERITAGE surveys of the
Magellanic Clouds associated with \hii\ regions that were classified by
\citet{p12} as candidate optically thick and thin objects.  

Since PAHs are easily destroyed by UV radiation, in principle, nebular optically
thin conditions may be confirmed by low peripheral PAH abundance.  We find that
the use of PAHs as a diagnostic for nebular conditions is compromised
by the strongly non-uniform spatial distribution of dust relative to the
ionized gas.  Nevertheless, for metallicities allowing significant
dust formation, as in the LMC, optically thick \hii\ regions clearly
show much higher 8~\mic\ surface brightness, with a median
value about 6 times higher than for optically thin objects.  The lower
8~\mic\ emission in optically thin objects is unlikely to be due to
lower heating rates since on average the stellar ionizing fluxes are
higher in optically thin objects \citep{p12}.  Thus, the
8~\mic\ emission can offer important supporting diagnostic data on
optical depth at higher metallicities.

In contrast to the LMC, we find no differentiation in the low PAH
levels seen in the optically thick and thin nebulae of the SMC.  These
results are consistent with the model of \citet{sandstrom10} in which
low-metallicity PAH abundance is regulated by low production rates in
molecular clouds and high destruction rates by stellar UV radiation.
This dominates the variations in PAH abundances of star-forming
regions and masks any differentiation due to optical depth effects.
Thus, at this much lower metallicity, it appears that PAHs are simply
too underabundant to serve as diagnostics for Lyman continuum opacity.

The very small dust grains that produce the 24 \mic\ emission are
more resilient to UV radiation and well known to correlate with
star-forming regions.  We confirm that it is associated with 
star formation, having more uniform morphological correspondence to luminous
\hii\ regions and star-forming knots.  We again find that the optically
thick \hii\ regions show a significant offset, a factor of about 3, in median
24~\mic\ surface brightness relative to the optically thin objects in
the LMC.  However, the offset here is due to the association
with denser gas in optically thick regions, rather than destruction in
optically thin regions.  As with the PAH emission, there is no
discernible difference with nebular optical depth in the SMC, again
attributable to low dust abundance.

Thus, we find that the low metallicity in the SMC apparently inhibits the
formation of PAHs and dust such that we cannot use the 8~\mic\ and
24~\mic\ emission as diagnostics of nebular radiative transfer.
This is further confirmed by inspection of the dust mass surface
densities, showing no significant difference between the optically
thick and thin objects in the SMC.  In contrast, the LMC shows that
the optically thick objects have higher median dust mass surface
density by a factor of 1.7 compared to the optically thin objects, and
the median GDR similarly is 1.8 times higher.
This contrast in PAH diagnostics is consistent with the suggestions
of a transition in ISM dust conditions at metallicities
just above the SMC value \citep{draine07,engelbracht05}, such that the
PAH contribution to dust mass drops precipitously in metal-poor
environments.  For our purposes, the decrease in 24~\mic-emitting hot
dust also precludes the use of this emission as a useful diagnostic of
nebular conditions in these environments. 

Hence, our findings suggest that at higher metallicities, the
8~\mic\ PAH and 24~\mic\ hot dust emission can offer useful
diagnostics of \hii\ region radiative transfer.  We do caution that
there is significant overlap in the distributions of properties
between the optically thick and thin objects.  Much of this
degeneracy is due to the fact that optical depth is not a binary
classification, but rather, a continuous quantity, and efforts to bin
objects into two categories will necessarily cause overlap in the
distributions.  We further caution that the optical depth
classifications of \citet{p12} have a large degree of subjectivity, as
do our reclassifications based on the 8\mic/[\oiii] maps in Section~2.
As stressed by \cite{p12}, IPM can only offer a first-order
estimate of optical depth for a single ratio map, and so 
classifications of individual objects should be regarded as tentative.
The 8~\mic\ and 24~\mic\ emission can therefore provide valuable
additional diagnostics when combined with the nebular emission-line ratio
maps.  As discussed in Section~2, since PAHs are more sensitive to UV
radiation than atomic species, they seem to be sensitive to a somewhat higher
optical depth threshold.

\acknowledgments

This work was supported by the National Science Foundation, grant
AST-1210285.  M.R. acknowledges support from CONICYT (Chile) through
FONDECYT grant No. 1140839 and partial support through project BASAL PFB-06.
We also thank the anonymous referee for helpful comments.

\bigskip\bigskip

\appendix

As described in Section~2, we classified all the MCELS objects as optically
thick or thin, based on the 8\mic/[\oiii] ratio map.  The
classifications were evaluated by J.L.-H.  Objects marked with 
asterisks indicate ones for which our classifications differ from those of
\citet{p12}, which were based on [\sii]/[\oiii] ratio maps. \\

Our LMC classifications are as follows. \\

Optically thick:
MCELS-L4, L6, L8, L11, L13*, L15*, L25, L28, L29*, L32, L33, L35, L47, L54, L60, L65*, L69, L70, L73, L78, L93, L95, L96, L108*, L125, L127, L130, L131, L132, L134, L135*, L136, L140, L143, L144, L149, L162, L173, L181, L188, L192, L193, L194, L197, L201*, L204, L206, L208, L212, L213, L215, L216, L218, L219, L222, L226, L227, L229, L230, L237, L238, L244, L251, L255*, L257, L261*, L264, L268, L274, L278, L285, L286, L290, L292, L304, L310, L311, L318, L320, L332, L334, L335, L336, L339, L340, L341, L342, L343*, L345, L348, L352*, L353*, L354, L355, L357, L369, L372, L374, L377, L382, L384, L385, L389, L390, L391, L393, L400. \\

Optically thin:
MCELS-L1, L2, L3, L5*, L9*, L10, L12, L14*, L16, L17*, L18*, L20, L21, L22*, L23*, L24, L27, L34*, L36*, L38, L39, L40, L42, L43*, L44, L45, L48, L49, L52*, L55*, L56, L58, L59, L61*, L63, L67, L71, L72*, L74*, L75*, L77, L79*, L80*, L86, L92, L97*, L98*, L99, L101, L102, L103, L104*, L106, L107*, L109*, L114*, L118*, L119, L121, L122*, L128*, L137, L138*, L141*, L146, L147, L148, L150, L151, L152, L155, L157*, L163, L165, L167*, L168, L169, L170*, L171, L174, L175*, L176, L177, L180, L182, L184, L191, L200, L202, L203*, L207*, L209, L210*, L211, L217*, L223, L231, L232, L239, L240, L241, L242, L248, L250*, L252*, L253, L254, L258, L259, L260, L267, L277*, L284*, L288*, L295, L300*, L302, L303, L305, L306, L307, L315*, L316*, L319*, L321*, L323, L325*, L326, L328, L333, L337*, L338, L344*, L346*, L347*, L351*, L356, L361*, L362*, L365, L367, L373, L379, L380*, L386, L394*, L395*, L396, L401*. \\

Our SMC classifications are as follows. \\

Optically thick:\\
MCELS-S1*, S4, S6*, S7*, S9, S10*, S14*, S27, S32, S33, S34, S42, S47, S71, S80, S81, S85, S86, S92, S93*, S96*, S97, S101, S104*, S105*, S107, S113*, S115, S119, S123, S126*, S131*, S132*, S139, S140*, S142, S143*, S149, S151*, S157*, S161, S162*, S164, S166, S167, S169, S170, S172*, S173, S175*, S176, S177*, S178, S179, S183*, S184, S185*, S187*, S188, S189, S192*, S196, S198, S204*, S206*, S208. \\

Optically thin:\\
MCELS-S2*, S3*, S5, S8*, S15*, S16, S17*, S18*, S19*, S20*, S22*, S23*, S24*, S25*, S26*, S28*, S29*, S30*, S31*, S35*, S36, S37, S38*, S39*, S40*, S43*, S44*, S45*, S46, S48*, S49*, S51*, S52*, S54*, S55*, S56*, S57*, S59*, S60*, S61, S62, S63, S64*, S65, S66, S67, S68*, S70*, S72*, S73, S74*, S77*, S78*, S79*, S82, S83*, S84*, S87*, S88*, S89*, S90*, S91*, S94, S95*, S98, S99, S102*, S103*, S106*, S108, S109, S110*, S111, S112*, S114, S116*, S117*, S121*, S124, S125, S127, S128*, S130*, S133*, S134, S135*, S137*, S138, S141*, S144*, S145, S146*, S147*, S148, S150*, S152, S153, S154, S155*, S156*, S158*, S159, S160, S168*, S171*, S174*, S180*, S181*, S182, S186*, S190*, S191*, S195*, S197*, S199, S200*, S207*, S209*, S210*, S211*, S212*, S213, S214*.

\end{document}